\documentclass[11pt,oneside]{article}
\usepackage{graphicx}
\begin{document}
 
\title{{\bf FORMATION OF GALACTIC SYSTEMS IN LIGHT OF THE MAGNESIUM ABUNDANCE
IN FIELD STARS.III.THE HALO}}
\author{{\bf V.~A.~Marsakov, T.~V.~Borkova}\\
Institute of Physics, Rostov State University,\\
194, Stachki street, Rostov-on-Don, Russia, 344090\\
e-mail: marsakov@ip.rsu.ru, borkova@ip.rsu.ru}
\date{accepted \ 2006, Astronomy Letters, Vol. 32 No. 8, P.545-556}
\maketitle

\begin {abstract}
Data from our compiled catalog of spectroscopically determined 
magnesium abundances in dwarfs and subgiants with accurate 
parallaxes are used to select Galactic halo stars according to 
kinematic criteria and to identify presumably accreted stars 
among them. Accreted stars are shown to constitute the majority 
in the Galactic halo. They came into the Galaxy from disrupted 
dwarf satellite galaxies. We analyze the relations between the 
relative magnesium abundances, metallicities, and Galactic 
orbital elements for protodisk and accreted halo stars. We 
show that the relative magnesium abundances in protodisk halo 
stars are virtually independent of metallicity and lie within 
a fairly narrow range while presumably accreted stars demonstrate 
a large spread in relative magnesium abundances up to negative 
[Mg/Fe]. This behavior of protodisk halo stars suggests that the 
interstellar matter in the early Galaxy mixed well at the halo 
formation phase. The mean metallicity of magnesium-poor 
($[Mg/Fe]<0.2$~dex) accreted stars has been found to be displaced 
toward the negative values when passing from stars with low 
azimuthal velocities ($| \Theta |<50$~km\,s$^{-1}$)to those with 
high ones at $\Delta [Fe/H]\approx 0.5$~dex. The mean apogalactic 
radii and inclinations of the orbits also increase with 
increasing absolute value of $|\Theta |$ while their 
eccentricities decrease. As a result negative radial and 
vertical gradients in relative magnesium abundances are observed
in the accreted halo in the absence of correlations between the 
[Mg/Fe] ratios and other orbital elements, while these 
correlations are found at a high significance level for 
genetically related Galactic stars. Based on the above 
properties of accreted stars and our additional arguments, we 
surmise that as the masses of dwarf galaxies decrease, the maximum 
SN\,II masses and hence, the yield of $\alpha$-elements in 
them also decrease. In this case, the relation between the 
[Mg/Fe] ratios and the inclinations and sizes of the orbits of 
accreted stars is in complete agreement with numerical 
simulations of dynamical processes during the interaction
of galaxies. Thus the behavior of the magnesium abundance in 
accreted stars suggests that the satellite galaxies are disrupted 
and lose their stars en masse only after dynamical friction 
reduces significantly the sizes of their orbits and drags them 
into the Galactic plane. Less massive satellite galaxies are 
disrupted even before their orbits change appreciably under 
tidal forces.
\\


\end {abstract}

\section*{Introduction}

A detailed analysis of the chemical composition of the 
metal-poorest field stars, together with their Galactic 
orbital elements, makes it possible to reconstruct the 
formation history of the halo in the early Galaxy. Metal-poor 
stars are currently believed to form at least two subsystems, 
an accreted halo and a protodisk halo. The first subsystem 
was formed by the stars and globular clusters captured by the
Galaxy at different times from debris of dwarf satellite
galaxies disrupted by its tidal forces. The stars of the 
second subsystem, along with the metal-richer Galactic stars, 
were formed mainly from the matter of a single protogalactic 
cloud. It is unlikely that the interstellar matter from 
which the stars of these two metal-poor subsystems were 
formed has experienced an exactly coincident chemical 
evolution. Therefore, it would be interesting to search 
for subtle differences between them that could shed light 
on the histories of star formation and matter mixing inside 
and outside the single protogalactic cloud. Owing to the 
favorable "geographic" position of the Sun in the Galactic
plane, we have an opportunity to observe the stars of
all its subsystems in the immediate vicinity of the Sun
and to analyze in detail their chemical composition.
Indeed even the stars that rise high above the Galactic 
plane cross this plane twice during one revolution around 
the Galactic center and, if the perigalactic radii of their 
orbits are smaller than the solar one they sooner or later 
come close to the Sun. In addition it is interesting to 
identify stars with identical current Galactic orbits 
(i.e., those that presumably have a common origin) in the 
accreted halo and to try to trace the pattern of variations 
in the relative abundance, e.g., of $\alpha$-elements with 
increasing metallicity for them. If a clearly traceable 
relation is found between these parameters this, on the one 
hand will serve as further evidence for their genetic 
relationship and on the other hand will make it possible 
to compare the star formation rates and the patterns of 
chemical evolution in this group and in the early Galaxy.

The stars that currently belong to the protodisk halo were 
formed over a fairly short period. This follows from the 
fact that the firmly established earliest traces of enrichment 
of the interstellar medium by type~Ia supernova (SN\,Ia) 
ejecta are observed only in the stars of the later formed 
thick-disk subsystem (Mashonkina and Gehren~2001; Marsakov 
and Borkova~2005). The evolution time of close binary stars 
that subsequently explode as SNe\,Ia is short, $0.5-1.5$~Gyr 
(see, e.g., Matteucci~2001; Tsujimoto et al.~1995). 
Exclusively higher-mass ($M>10 M_\odot$) stars exploding as 
type II supernovae (SNe\,II) are currently believed to have 
enriched the interstellar medium with heavy elements at 
earlier stages. Their characteristic evolution time is only 
$\approx 30$~Myr. Almost all of the nuclei of $\alpha$-elements 
are formed in SNe\,II. Concurrently a small amount of iron-peak
elements is also produced in these SNe while the bulk of them 
is ejected into the interstellar space during SN\,Ia explosions. 
Calculations show that the yield of various elements in SNe\,II 
depends on several parameters. In particular the yield of the 
socalled primary $\alpha$-elements (O, Ne, and Mg) being 
synthesized in hydrostatic processes in a carbon core and 
explosive Ne/C burning in a presupernova's shell sources 
depends strongly on the stellar mass (see, e.g., Thielemann 
et al.~1996; Nomoto et al.~1997). In contrast, the amount of 
iron-group elements depends on an explosion mechanism 
related to the iron core size. Therefore, the relative 
abundances of the primary $\alpha$-elements ([$\alpha$/Fe]) 
in the ejecta of different SNe\,II even with the same mass can 
differ markedly. Fortunately the next generation of stars is 
born in an interstellar medium that was generally enriched by
several supernovae and mixing partially levels off the 
dependence of the yield on their explosion parameters.
Hence, the variations in the upper boundary of the initial 
mass function for stars that exploded at different times 
inside and outside the early Galaxy can be estimated from 
the relative abundances of various elements in genetically 
related stars. Concurrently because of the difference between 
the evolution times of SNe\,II and SNe\,I we can try to trace 
the star formation rate for this stellar ensemble by the 
coordinates of the characteristic knee in its [$\alpha$/Fe]--[Fe/H]
diagram toward the sharp decrease in the relative abundance 
of the primary $\alpha$-elements with increasing total 
heavy-element abundance at the onset of SNe\,Ia explosions, 
i.e., $\sim 1$~Gyr later.

The best-studied primary $\alpha$-elements are oxygen and 
magnesium because they exhibit several absorption lines in 
the visible spectral range. Although the theory predicts a 
dependence of the Mg yield (in contrast to oxygen) on mixing 
parameters in the presupernova's atmosphere, we preferred to 
use here precisely this element, since its abundance in stars 
is determined much more reliably than that of oxygen.
For the analysis, we took data from our compiled catalog of 
spectroscopically determined magnesium abundances (Borkova 
and Marsakov~2005). Almost all of the magnesium abundances 
in dwarfs and subgiants in the solar neighborhood determined 
by synthetic modeling of high-dispersion spectra published
before January~2004 were gathered in this catalog. This catalog 
is several times larger than any homogeneous sample that has 
been used until now to analyze the Galactic chemical evolution. 
The relative magnesium abundances in the catalog were derived 
from 1412~spectroscopic determinations in 31~publications for 
867~dwarfs and subgiants using a three-pass iterative averaging 
procedure with a weight assigned to each primary source and each
individual determination. The internal accuracy of the 
catalogued relative magnesium abundances for metal-poor 
([Fe/H]$<-1.0$~dex) stars is $\varepsilon$[Mg/Fe]=$\pm 0.07$~dex. 
The metallicities for the stars were obtained by averaging about 
2000~determinations from 80~publications; the accuracy for 
metal-poor stars was estimated to be $\varepsilon$[Fe/H]=$\pm 0.13$~dex.
The distances to the stars and their space velocities were 
calculated on the basis of data from currently available 
high-precision catalogs. We used trigonometric parallaxes with 
errors smaller than 25\,\% and, if these were lacking, photometric 
distances calculated from $uvbyH_\beta$ photometry. Based on a 
multicomponent model of the Galaxy (Allen and Santillan~1991)
containing a disk a bulge, and an extended massive halo, we 
calculated the Galactic orbital elements by simulating 
30~revolutions of a star around the Galactic center. The 
Galactocentric distance of the Sun was assumed to be 8.5~kpc, 
the rotational velocity of the Galaxy at the solar 
Galactocentric distance was 220~km\,s$^{-1}$ and the 
velocity of the Sun with respect to the local standard of 
rest was $(U_\odot, V_\odot, W_\odot )= (-11, 14, 7.5)$~km\,s$^{-1}$ 
(Ratnatunga et al.~1989).

Previously we analyzed the relative abundances of magnesium and 
europium (an r-process element) in 77~nearby stars, of which 
only 23~stars belong to the Galactic halo (Borkova and 
Marsakov~2004). All of the presumably accreted stars that we 
identi ed according to kinematic criteria were found to exhibit 
[Mg/Eu] ratios that differ sharply from those in the stars 
genetically related to the Galaxy. (These differences serve as 
a further argument in favor of their extragalactic origin.) 
Analysis of the deviations led us to conclude that the 
maximum masses of the very first stars exploded as SNe\,II 
outside the Galaxy were much lower than those inside it. 
In contrast, the supernovae exploded outside the Galaxy 
somewhat later turned out to be more massive than those in 
the Galaxy itself. This intermediate [Mg/Eu] ratio in the
metal-poor stars of the early Galaxy may be indicative
of more intense mixing inside the single protogalactic cloud.

This paper completes our systematic description of the 
chemical and spatial--kinematic properties of the stars that 
are currently in the solar neighborhood but that belong to 
different Galactic subsystems based on data from our compiled 
catalog. In our previous papers, we considered the properties 
of thick-and thin-disk stars (see Marsakov and Borkova~(2005 and
2006), respectively).

\section*{IDENTIFICATION OF STARS OF THE HALO SUBSYSTEMS}

In our paper devoted to the thick disk, we justified the choice of 
the residual stellar velocity relative to the local standard of 
rest res $V_{res}=175$~km\,s$^{-1}$ as a criterion for separating the 
thick-disk and halo stars. We optimized the specific value of 
this criterion by minimizing the number of metal-rich 
([Fe/H]$>-1.0$) stars of our sample in the identified halo 
subsystem and metal-poor stars in the thick disk. The need for 
this optimization procedure is dictated by the presence of
adistinctde cit of high-velocity ($V_{res} >100$~km\,s$^{-1}$)
stars near this point in their metallicity distribution
suggesting the absence of a smooth transition between the halo 
and the thick disk (see Marsakov and Suchkov~1977).

In identifying the stars of an extragalactic origin (which were 
called here accreted stars), we assumed that the stars born in 
a monotonically collapsing single protogalactic cloud could 
not be in retrograde orbits. In this case, all of the stars with 
retrograde orbits around the Galactic center may be considered
accreted ones. However as can be seen from the spatial 
distribution of presumably accreted globular clusters (see 
Fig.~6 in Borkova and Marsakov~(2000)), not only the orbits of 
their parent satellite galaxies differed significantly before 
their disruption. Therefore we included all of the stars with 
the same high residual velocities as those for the retrograde 
ones, i.e., $V_{res} >240$~km\,s$^{-1}$ in the group of presumably
accreted stars. As we see from Fig.~1a, it is at this critical 
residual velocity that stars in retrograde orbits appear in our 
sample. We also included two more slower stars (HD~103723 and 
HD~105004)is this group. (As will be shown below they both 
simultaneously exhibit high metallicities ([Fe/H]$\approx -0.73$) 
and anomalously low magnesium abundances ([Mg/Fe]$<0.1$), which
are a typical of the stars of the protodisk halo where they fell 
according to the residual velocity criterion: $V_{res}=211$ and 
217~km\,s$^{-1}$ respectively.) Figure~1\,b shows that the spread 
in apogalactic radii of the stellar orbits increases sharply as 
one passes through the critical residual velocity; the orbits for 
stars with prograde motions have even systematically larger sizes
than those for stars with the same residual velocity relative to 
the local standard of rest, but with retrograde motions.

We emphasize that, based only on a kinematic criterion, we cannot 
identify the stars of a particular Galactic subsystem with absolute 
confidence. However since the main goal of this paper is to study 
the abundance differences between some of the chemical elements in 
stars of the two halo subsystems, we decided to use this 
chemical-composition independent parameter lest an artificial selection 
be introduced. Therefore, the personal identification of each star 
should be considered to be preliminary and to require a further 
refinement. However this criterion is quite appropriate in 
statistical analyses to reveal any relations between chemical 
composition and other parameters of stars of different Galactic
subsystems.

\section*{RELATION BETWEEN THE IRON AND MAGNESIUM ABUNDANCES}

Thus our kinematic criteria identified a total of about 20~protodisk 
halo stars, while more than a hundred stars from our sample belong 
to the accreted halo. By the "protodisk" halo we mean the subsystem
that consists of stellar objects that are the first to be formed from 
the matter of a collapsing single protogalactic cloud and that are 
genetically related to the stars of all the subsequent generations. 
The appreciably increasing number of genetically related stars with 
metallicity (see Fig.~2) suggests that more active star formation 
in the early Galaxy began slightly later when the mean metallicity 
of the interstellar medium reached [Fe/H]$\approx -1.0$~dex, while 
the "accreted" objects subsequently formed the bulk of the Galactic
halo. By this term we mean all of the objects that were born outside 
the single protogalactic cloud, i.e., in the nearest satellite 
galaxies or in isolated protogalactic fragments, and that subsequently 
escaped from them under the Galactic tidal forces.

Figure~2 shows the metallicity--relative magnesium abundance diagram 
for our catalog, where different symbols denote stars belonging to the 
accreted halo, to the protodisk halo, and to both disk subsystems. We see 
from the figure that the relative magnesium abundances in protodisk halo 
stars are virtually independent of metallicity and that all stars of 
this halo lie above the dashed line drawn through [Mg/Fe]=0.2~dex i.e.,
the relative magnesium abundances in protodisk halo stars are higher 
than those in the bulk of the thin-disk stars (Marsakov and Borkova~2005).
One cannot but notice the presence of an appreciable number (five) of 
relatively metal-rich ([Fe/H ]$>-1.4$~dex) stars with high relative 
magnesium abundances ([Mg/Fe]$>$0.5~dex) in the diagram; no theoretical 
models of chemical evolution predict their appearance. The impression 
is that when a metallicity [Fe/H]$\approx -1.5$~dex is reached in the 
interstellar medium more massive SNe\,II begin to explode in it.
Given the lack of data on such stars in our sample, we will not consider 
this point here. The remaining protodisk halo stars lie in a narrow 
strip in the diagram in good agreement with the theoretically predicted 
dependence (see, e.g., the model by Matteucci and Francois~(1989)). This 
behavior of protodisk halo stars suggests that, at least in the initial 
stage of its formation the interstellar matter in the early Galaxy 
either was well mixed or SNe\,II of the same mass exploded in all 
local volumes. In contrast, the presumably accreted stars exhibit a 
large spread in relative magnesium abundances in Fig.~2 that extends
to negative [Mg/Fe]. The anomalously low relative magnesium abundances 
in some of the accreted stars are usually explained by an extremely 
low star formation rate in the dwarf satellite galaxies where these
stars were born (Gilmore and Wise~1998). However our analysis of the 
relative magnesium and europium abundances in a small sample of nearby 
field stars showed that all of the presumably accreted stars exhibited 
an [Eu/Mg] ratio that differed sharply from its Galactic value (Borkova 
and Marsakov~2004). Since the relative yield of these elements depends 
solely on the masses of the SN\,II progenitor stars where they are 
synthesized, we believe that a more likely mechanism of the magnesium 
abundance variations in accreted stars is the difference between the 
initial mass functions in their parent dwarf satellite galaxies.
Therefore, it is interesting to try to identify genetically related 
stars in the accreted halo.

In the above paper we drew attention to a small group of stars with 
identical azimuthal velocities and pointed out that their orbital 
elements are in satisfactory agreement with those of the most massive
Galactic globular cluster $\omega$~Centauri which is believed to have 
been the nucleus of a dwarf galaxy in the past (Larson~1996). Shortly 
after the appearance of our paper Meza et al.~(2005) published their
paper in which the authors used numerical simulations from Abadi et 
al.~(2003) to investigate the orbital characteristic of dwarf 
satellite galaxies whose orbits were reduced significantly and dragged 
into the Galactic plane by dynamical friction even before their 
disruption. Disrupted on very eccentric orbits coplanar with the 
disc, such galaxies are expected to shed stars in "trails" of distinct 
orbital energy and angular momentum during each pericentric passage. 
Therefore, to an observer located between the pericentre and apocentre 
of such orbits, these trails would show as distinct groupings of stars
with low vertical velocities and a broad, symmetric, often double-peaked 
distribution of Galactocentric radial velocities. Based on the 
recommendations of Meza et al.~(2005), we identified stars that were 
lost by the dwarf galaxy whose center was the cluster $\omega$~Cen by 
the azimuthal and vertical velocities in the ranges $-50 \leq \Theta \leq 0$ 
and $|W|<65$~km\,s$^{-1}$, respectively. There are 18~such stars in 
our sample. In the [Mg/Fe]--[Fe/H] diagram (see Fig.~2), all of them 
(except one deviated star G\,233--026) lie along a narrow strip that 
exhibits an approximately constant ratio of [Mg/Fe]$\approx 0.35$~dex 
at low metallicities ([Fe/H]$\leq -1.3$~dex). Subsequently however as 
the metallicity increases, the relative magnesium abundance decreases 
sharply to negative values starting from this point. This behavior 
closely resembles the expected [Mg/Fe]--[Fe/H] relation derived in a 
closed model of chemical evolution (see, e.g., Matteucci and Francois~1989),
which is independent evidence for the genetic relationship between the 
identified stars. (Meza et al.~(2005) identified 11 members of this
group and also pointed out the existence of a genetic relationship 
between them.) Hence, the low relative magnesium abundances in the 
metal-richest stars of this group resulted from the SN\,Ia explosions 
that began in their parent protogalactic cloud and that ejected a 
large number of iron atoms into the interstellar medium and reduced the 
[Mg/Fe] ratio. The considerably lower metallicity of the knee point 
in this diagram than that in the Galaxy suggests that the stars of 
the Centaurus moving group were formed from matter in which the 
star formation rate was considerably lower than that in the early 
Galaxy. Hence, at least in this, presumably initially massive 
($M\approx 10^9 M_\odot $; see Tshuchiya et al~.2003) disrupted 
satellite galaxy the mean masses of the SN\,II progenitor stars
were the same as those in our Galaxy.

Let us divide all our presumably accreted stars into four azimuthal 
velocity ranges by $\Theta$=(-250, -50, and +50)~km\,s$^{-1}$. In 
the ($\Theta$--[Mg/Fe]) and ([Fe/H]--[Mg/Fe]) diagrams, the stars 
of each range are plotted by different symbols (see Figs.~3\,a and 3\,b,
respectively). The dashed lines separate the relatively magnesium-poor 
and magnesium-rich stars by the ratio dex. The overwhelming majority 
(12 of 16) magnesium-poor accreted stars of moderate metallicity 
(-1.3~dex$<$[Fe/H]$<$-0.7~dex) fell within the range 
$|\Theta |<50$~km\,s$^{-1}$ (the filled triangles in the diagram). 
(Note that there are also stars among these 12~stars that we 
attributed to the accreted halo, despite their residual velocities 
that did not reach the critical value; their numbers are given in 
Fig.~3b.) A test showed that the orbits of these 12~stars go away 
not very far from the Galactic plane: together with the four stars 
from the Centaurus group located here 
$\langle|Z_{max}|\rangle = 8.0\pm 1.2$~kpc for them. At the same time, 
the mean largest distance of the orbital points from the Galactic 
plane for the remaining five stars (see Fig.~3\,b) that slowly 
revolve around the Galactic center but that are metal-poorer ones 
([Fe/H]$<-1.3$~dex) is more than twice as large: $19.1 \pm 2.9$~kpc.
Thus almost all of the magnesium-poor stars of moderate metallicity 
are found to fall within the solar neighborhood from the satellite 
galaxies that began to lose their stars only after their orbits were 
strongly distorted by the gravitational interaction with the Galactic 
disk and were brought closer to its plane. Their masses must 
probably be high enough to resist the Galactic gravitational forces 
for a long time and to retain stars near them. The assumption that
the masses of these satellites are fairly high is also con firmed by 
the fact that their chemical evolution is well interpreted in terms 
of the closed model. (Indeed, only at a mass of the protocloud above 
a certain limit is the enriched gas not swept out from it by
supernova explosions.) In contrast, the disruption of less massive 
satellites began earlier; therefore, the orbits of the stars lost by 
them are generally higher. The low ratios [Mg/Fe]$<$0.2~dex at 
metallicities in the range -1.4~dex$<$[Fe/H]$<$-2.4~dex for the
above ve slow but high-orbit magnesium-poor stars can hardly be 
explained by the low star formation rate in their parent dwarf 
galaxies alone. Here, a deficit (or complete absence) of massive 
($M>20 M_\odot$) SNe\,II which are the main suppliers of primary 
$\alpha$-element atoms, is most likely responsible for the low 
magnesium abundance.

Stars with large absolute values of the azimuthal velocities in 
the solar neighborhood also have large apogalactic radii of their 
orbits. Therefore, according to Abadi et al.~(2003), they must come 
into the Galaxy predominantly from less massive and, hence, early 
disrupted dwarf galaxies in which there seems to have been no star 
formation after the first generation SNe\,Ia explosions. Let us examine 
the behavior of stars from the higher-velocity $\Theta$ ranges
in the [Fe/H]--[Mg/Fe] diagram. The stars from symmetric (in 
velocities), but different (in direction of rotation) ranges 
($-250 < \Theta < -50$~km\,s$^{-1}$ and $\Theta >50$~km\,s$^{-1}$) 
are randomly distributed over the field of the diagram and reveal no 
clearly traceable sequences that would be indicative of a genetic
relationship between some of them. However it may be noted that 
positively rotating stars are, on average, metal-richer and 
magnesium-poorer ($\langle$[Fe/H]$\rangle=-1.47$ and $-1.91$~dex, 
while $\langle$[Mg/Fe]$\rangle=0.23$ and 0.34~dex respectively for 
positively and negatively rotating groups of stars). At the same 
time, all seven stars with the highest negative azimuthal velocities 
(the crosses in Fig.~3\,b) form a compact group with 
$\langle$[Fe/H]$\rangle=-1.68$~dex and a small spread, 
$\sigma_{[Fe/H]}=0.14$~dex which is almost equal to the error in 
metallicity and show anomalously low relative magnesium abundances 
($\langle$[Mg/Fe]$\rangle=0.09 \pm 0.02$~dex). Thus there are no 
more or less numerous groups with a common origin among the rapidly 
rotating accreted stars; therefore, all of them seem to have come 
into the solar neighborhood from various stellar ensembles.
The most noticeable common property that we can point out for 
magnesium-poor stars so far is the displacement of the mean 
metallicity to the negative values for them when passing from 
the stars that revolve slowly around the Galactic center to those
that revolve rapidly: $\langle$[Fe/H]$\rangle =-1.2 \pm 0.1$~dex 
and $-1.7 \pm 0.1$~dex, respectively for the group with 
$|\Theta |<50$~km\,s$^{-1}$ and $\Theta  < -250$~km\,s$^{-1}$.

Note that two thirds of our stars in the velocity range 
$|\Theta |>50$~km\,s$^{-1}$ have only one published relative 
magnesium abundance determination while within the range 
$|\Theta |<50$~km\,s$^{-1}$ the [Mg/Fe] ratios for half of 
the stars were averaged over several sources. As a result, 
more than one source is available only for $\approx 10$\,\% 
of the metal-poor ([Fe/H]$<-1.0$~dex) magnesium poor 
([Mg/Fe]$<0.2$~dex) stars. Therefore, it is desirable to confirm 
the results of this section by further observations.

\section*{RELATION BETWEEN CHEMICAL COMPOSITION AND ORBITAL ELEMENTS OF STARS}

Let us analyze in more detail the relations between the chemical 
composition of accreted stars and their orbital elements. 
Figure~4\,a where only accreted halo stars are plotted in 
the $\Theta$--[Mg/Fe] diagram shows more clearly the trend found 
in the previous section. We see that the lower limit for the relative 
magnesium abundance in accreted stars is almost independent of their 
azimuthal velocity. Since the right and left corners of the 
diagrams are totally devoid of stars, the upper envelope of the 
diagram is naturally described by two inclined lines that 
intersect near $\Theta \approx 0$~km\,s$^{-1}$. In other words 
the upper limit for the [Mg/Fe] ratio in accreted stars 
systematically decreases with increasing absolute value of the 
azimuthal velocity. As a result, we see that accreted stars 
with high relative magnesium abundances tend to have low orbital 
velocities around the Galactic center at the solar Galactocentric 
distance. At the same time, stars with low [Mg/Fe] ratios are 
distributed more uniformly in all permitted values.

Since the linear velocity of a star is known to depend its orbital 
phase, it is not surprising that the $\zeta$--[Mg/Fe] diagram in 
Fig.~4\,b where the orbital inclination is defined as 
$\zeta =\arcsin Z_{max}/Ra$ and its sign is the same as that of 
the azimuthal velocity also shows a similar structure. In this 
diagram, all points can also be bounded by three straight lines
that form a triangle. Here, however a straight line represents 
well the lower envelope only in the segment $|\zeta|\leq 45^\circ$  
while there are no stars with such low magnesium abundances 
outside this range; the stars concentrate to $\zeta =0^\circ$ at 
any [Mg/Fe]. Thus the stars with orbits lying near the Galactic 
plane are found, first, to constitute the majority of accreted
stars and, second, to exhibit a wider range of relative magnesium 
abundances in both directions.

In Figs.~5\,a and 5\,b, the [Mg/Fe] ratios are plotted against the 
maximum distances of the orbital points of stars from the 
Galactic center and plane. Both diagrams shows highly significant 
correlations for accreted stars: at the same correlation coefficients 
($r = 0.3 \pm 0.1$), the probability of chance occurrence of both 
correlations for the same number ($N$) of objects is $P_N\ll 1$\,\%.
(To ensure the reliability of the results, we discarded the five 
most distant points in the diagrams that often determine the 
correlation itself.) The two gradients in relative magnesium 
abundances were found to be negative rather than positive, as
for all of the genetically related stars in the Galaxy and nonzero 
above the $2.5\,\sigma$ level. (Although this is not seen on the 
scale of Fig.~5\,b, all of the accreted stars in our sample 
with maximum distances of their orbital points from the Galactic 
plane less than 1~kpc (13~stars) have very high relative magnesium 
abundances, [Mg/Fe]$>$0.3~dex.) The detected gradients are not 
related to the evolution of the Galaxy but only reflect the sizes 
of the orbits of the satellite galaxies, lying on which they lose 
their stars under the Galactic tidal forces. The sizes of the 
orbits for accreted stars and, hence, for their disrupted parent 
galaxies are found to increase, on average, with decreasing relative
magnesium abundance in them. Note, however that the lower envelopes 
in the diagrams more likely have an opposite slope and the magnesium 
abundance even slightly increases with both $Ra$ and $Z_{max}$. In
other words despite the higher on average, relative magnesium 
abundances in them, there are also the magnesium-poorest accreted 
stars among the stars with small sizes of their orbits.

We see from Figs.~5\,c and 5\,d that the same stars exhibit neither 
radial nor vertical metallicity gradients. However almost all of 
the stars here also occupy triangular regions in the diagrams that,
in addition to the vertical axis, are bounded by two straight lines 
(the dashed lines in the diagrams); the upper and lower envelopes 
have opposite slopes and converge near [Fe/H]$\approx -1.5$~dex. As 
a result, the mean metallicity at any distance from the Galactic
center and plane remains approximately constant, showing zero 
gradients, while the spread in [Fe/H] decreases with increasing 
distance. As with magnesium there are both metal-poorest and 
metal-richest accreted stars among the stars with the smallest sizes
of their orbits. It can be seen from a comparison of the two upper 
and two lower diagrams in Fig.~5 that a nearly solar magnesium 
abundance and a moderate (for old stars) heavy-element underabundance
([Fe/H]$\approx -1.5$~dex) are characteristic of the most distant orbits.

Protodisk halo stars are also plotted in Fig.~5. Our sample includes 
only 22~such stars, i.e., considerably fewer than the accreted stars 
(108~stars). Since such a small number does not allow statistically 
significant conclusions about the gradients in this subsystem to be 
drawn, we do not consider in detail the properties of its stars in 
this paper.

In Fig.~6, magnesium and iron abundances are plotted against 
eccentricities and perigalactic radii of the orbits for stars of 
different Galactic subsystems. In all diagrams, the stars of the 
genetically related subsystems (the thin and thick disks and the 
protodisk halo) form clearly traceable sequences. The sequences 
that intersect with the sequences of genetically related stars at a 
certain angle can also be traced for accreted stars (the corresponding 
regression lines are drawn). All correlations for genetically related 
stars are highly significant ($r \approx 0.60 \pm 0.02$ at $P_N\ll 1$\,\%).
In contrast, for accreted stars, only the correlations between their 
metallicity and orbital elements are highly significant (in both cases,
$r = 0.3 \pm 0.1$ at $P_N\ll 1$\,\%), while the mean relative 
magabundance in them is essentially uncorrelated with these orbital 
elements. Stars of different genesis are grouped in the opposite 
parts of the diagrams; exactly one half of the accreted stars have 
very high eccentricities, $e>0.9$ and small perigalactic radii of their 
orbits, $Rp <1$~kpc. The mean apogalactic radii their orbits are 
also small ($16 \pm 1$~kpc). The stars of the Centaurus moving group 
that came into our Galaxy from a disrupted, fairly massive dwarf galaxy
have precisely such orbital elements. This is understandable: more 
massive disrupted systems leave a more distinct trace among the Galactic 
field stars.

\section*{THE GALACTIC HALO POPULATIONS}

Thus the trends found above make it possible to present the following 
formation picture of the Galactic halo. The small relative number of 
protodisk halo stars suggests that the star formation rate in a 
contracting protogalactic cloud was considerably lower than that at 
later stages of its evolution. Nevertheless, the star formation was 
fairly intense, since the metallicity of the interstellar medium had 
managed to increase to [Fe/H]$\approx -1.0$~dex solely through SN\,II
ejections before SN\,Ia explosions began. The comparatively small 
spread in relative magnesium abundances in the stars of this subsystem 
indicates that the interstellar medium was well mixed in this 
evolutionary period of the Galaxy. Accreted stars constitute the bulk 
of the Galactic halo. They came into the Galaxy from the nearest dwarf 
satellite galaxies. According to the numerical simulations of the 
heir of archical Galactic halo formation performed by Abadi et al.~(2003),
massive satellite galaxies are disrupted up and lose their stars en 
masse only after dynamical friction reduces significantly the sizes of 
their orbits and brings them virtually into the Galactic plane. In 
contrast, less massive satellite galaxies (or those with orbits almost 
perpendicular to the disk) are disrupted even before their orbits 
change appreciably under the Galactic tidal forces. As a result, the 
stars lost by galaxies of different masses must have different kinematic 
characteristics. The relations between the iron and magnesium abundances 
and orbital elements of presumably accreted stars found here confirm this
hypothesis put forward theoretically solely from dynamical considerations.
Indeed, the 18~stars of the socalled Eggen moving group that we identified 
according to kinematic criteria exhibit a [Fe/H]--[Mg/Fe] sequence 
characteristic of genetically related stars. With a high probability 
these stars belonged in the past to a fairly massive satellite galaxy 
whose central nucleus was the globular cluster $\omega$~Cen. The 
position of the knee in relative magnesium abundance at 
[Fe/H]$\approx -1.3$~dex indicates that the star formation rate in 
this dwarf galaxy was lower than that in our Galaxy. The star formation 
in this galaxy continued so long that its metal-richest stars reached 
[Mg/Fe]$<$0.0~dex, i.e., a ratio that is even smaller than that for 
the Sun. (A lower star formation rate than that in the Galaxy may 
also have played a certain role.) However the low value of the maximum 
metallicity in the stars of the group (only [Fe/H]$\approx -0.7$~dex)
points to the cessation of further star formation in the parent galaxy.
The very small scatter of points in this sequence suggests that the 
interstellar medium is well mixed in it. At the same time, the mean 
ratio $\langle$[Mg/Fe]$\rangle \approx 0.35$~dex in the metal-poor 
stars of this group, which is equal to its Galactic value suggests 
that the mean SN\,II masses are comparable in both galaxies, since 
the yield of $\alpha$-elements depends mainly on the presupernova 
mass. In other words, the chemical composition of the stars in this 
former galaxy (and their orbital elements)indicate that it actually 
had evolved for a fairly long time (but less than our Galaxy) before 
disruption and the initial stellar mass function in it was the same 
as that in our Galaxy. The accreted stars of the sample with low 
azimuthal velocities $\Theta$ exhibit a similar pattern of relation 
between [Fe/H] and [Mg/Fe]: all metal-poor stars have relative 
magnesium abundances higher than those in several comparatively 
metal-rich stars for which the [Mg/Fe] ratio reaches its solar value. 
The orbits of these stars proved to be smallest and most eccentric,
as for the stars of the Centaurus group.

There is a considerable number of stars that are simultaneously poor 
in heavy elements ([Fe/H ]$<-1.0$~dex) and in magnesium 
([Mg/Fe]$<$0.2~dex) among the stars revolving around the Galactic 
center with higher velocities (both positive and negative), while 
all of the stars with the highest negative velocities have such 
anomalously low relative magnesium abundances for very metal-poor 
([Fe/H]$\approx -1.7$~dex) stars. The orbits of most stars with 
$|\Theta |>50$~km\,s$^{-1}$ are generally less eccentric and lie 
almost completely beyond the solar Galactocentric distance. Such orbits 
are characteristic of early disrupted low mass dwarf satellite 
galaxies. Shortly after their formation such low-mass galaxies begin 
to lose not only stars, but also interstellar gas while crossing the 
Galactic plane. In view of the loss of interstellar matter the star 
formation in them ends fairly rapidly; therefore, we barely see 
any metal-rich stars among them and it is unlikely that the anomalously 
low [Mg/Fe] ratios for such metal-poor accreted stars resulted 
solely from an extremely low star formation rate in their parent 
dwarf galaxies. Probably the main reason for this is that the 
initial stellar mass function in less massive dwarf galaxies is 
just truncated at the high masses. As a result, SNe\,II eject a 
smaller amount of light $\alpha$-elements into the interstellar
medium and the [Mg/Fe] ratios for the stars become anomalously 
low compared the stars of the same metallicity that are 
genetically related to the single protogalactic cloud.

As has already been noted above, we drew the same conclusion 
from the fact that the [Eu/Mg] ratios for a significant fraction 
of the accreted stars from the then analyzed small sample are 
considerably larger than those for the Galactic stars. Moreover in 
their recently published paper Decauwer et al.~(2005) explained 
the difference in the slopes of the correlations between the 
primary and secondary $\alpha$-elements in 21~mildly metal-poor 
stars by the fact that the stars exhibiting lower than average 
[$\alpha$/Fe] ratios are formed in low-mass protogalactic 
fragments that are unable to sustain the formation of very 
massive stars. The authors believe that even at a constant 
initial mass function low-mass clouds have a lower probability 
of forming very massive stars. Direct $\alpha$-element abundance 
determinations in stars of isolated dwarf galaxies also lead to 
a similar conclusion. Thus, for example, Tolstoy et al.~(2003) 
investigated the chemical composition in 15~red giants from four 
nearby low-mass ($\approx 2\times 10^7 M_\odot$), very diffuse 
dwarf galaxies without central nuclei (Sculptor Fornax, Carina
and Leo\,I). Almost all of the stars (14 of 15) fell on the 
([Fe/H]--[$\alpha$/Fe]) diagram in such a way as if they were 
born in a single protogalactic cloud, i.e., forming a rather 
narrow sequence that sinks from the point with coordinates 
([Fe/H]$-\approx 2.0$~dex and [$\alpha$/Fe]$\approx 0.3$~dex) 
to the point with ($-1.0$~dex and $-0.2$~dex). Not counting 
the two magnesium poorest stars (which is quite reasonable, 
bearing in mind the low accuracy of determining the elemental
abundances in giants), the remaining 20~stars fell within the 
region occupied by our accreted stars with high orbital 
velocities around the Galactic center. The authors suggest 
that such low relative $\alpha$-element abundances could 
result only in the case of low star formation rates and 
the absence of any contribution from very massive 
($>15--20 M_\odot$)stars to the chemical evolution.

The results of our previous paper (Borkova and
Marsakov~2002), where we found an anticorrelation between 
the Galactocentric distances and masses of globular clusters 
of a presumably extragalactic origin can also serve as a 
further argument in favor of the above suggestion. As a result, 
here we can also discern a statistical trend: predominantly 
low-mass metal-poor globular clusters turn out to be in distant
orbits. It appears that low-mass globular clusters are generally 
formed in low-mass dwarf galaxies. 

Thus the five independent facts organically fit within the 
framework of a single hypothesis. According to this hypothesis, 
metal-poor stars with anomalously low $\alpha$-element 
abundances come into our Galaxy from debris of low-mass 
satellite galaxies in which the chemical evolution proceeded 
not only slowly but also in the absence of massive SNe\,II.

Note that all our conclusions were formulated on the basis of 
statistical analyses and are not absolute in nature. We can 
also give examples of observed deviations from the trend found. 
Thus for example, several low-mass globular clusters (Ter~7, 
Ter~8, Arp~2, Pal~12) with anomalously low $\alpha$-element 
abundances ([$\alpha$/Fe]$<0.1$~dex) were formed in the 
currently disrupted very massive ($5 \times 10^9 M_\odot$; see 
Ibata and Rasoumov~1998) Sagittarius dwarf spherical galaxy to
which at least 10~globular clusters belonged earlier (see 
Borkova and Marsakov~(2004) and references therein). Since 
this galaxy is in a nearly polar orbit (Ibata et al.~1997), 
the globular clusters that escaped from it do not lie in 
the Galactic plane. (However according to Abadi et al.~(2003), 
such an orbit would not experience the disturbing in uence of 
Galactic tidal forces.)

The abundances of the remaining $\alpha$-elements as well as 
$r$- and $s$-process elements in the sample stars must be 
analyzed to reconstruct more accurately the formation 
history of the Galactic halo and to estimate the ratio of 
the contributions from slowed-down star formation and 
truncation of the upper limit for the stellar masses in 
low-mass dwarf galaxies that lead to the birth of stars 
with anomalously low relative magnesium abundances in them.

{\bf ACKNOWLEDGMENTS}:
We wish to thank the anonymous referee who found several 
inaccuracies in the manuscript. This work was supported 
in part by the Ministry of Science of Russia (project 
no.02.438.11.7001) and the Federal Agency for Education 
(projects RNP 2.1.1.3483 and RNP 2.2.3.1.3950).

\newpage

\begin{figure*}
\centering
\includegraphics[angle=0,width=0.80\textwidth,clip]{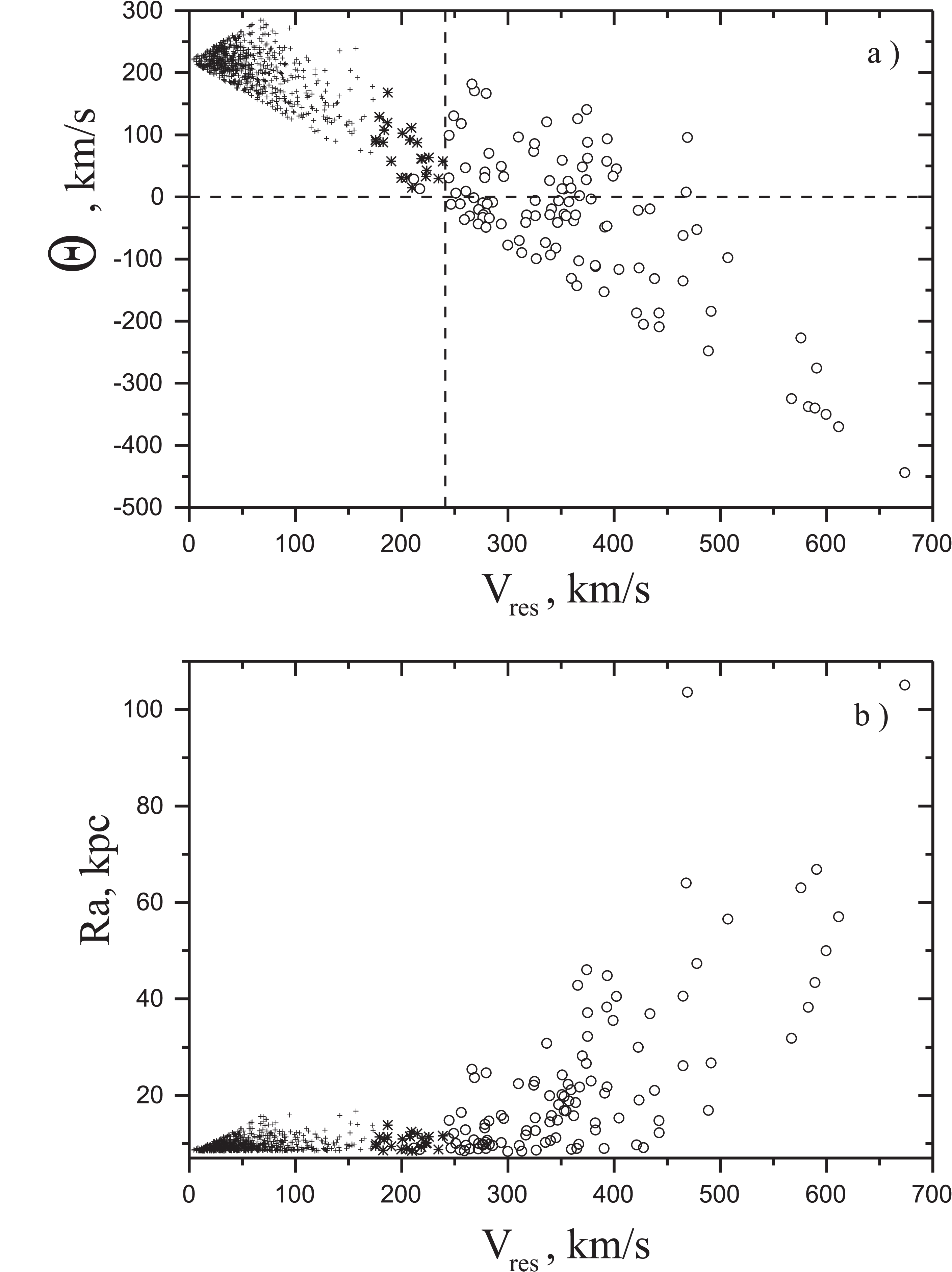}
\caption{Residual stellar velocities with respect to the local 
standard of rest vs.(a) azimuthal velocities and (b) apogalactic 
orbital radii: the crosses, asterisks, and circles indicate 
thin-and thick-disk stars, protodisk halo stars, and presumably 
accreted stars, respectively. The horizontal dashed line separates 
the stars with prograde and retrograde orbits around the Galactic 
center while the vertical line separates the accreted stars from 
the stars genetically related to the Galaxy.
}
\label{fig1}
\end{figure*}

\newpage

\begin{figure*}
\centering
\includegraphics[angle=0,width=0.84\textwidth,clip]{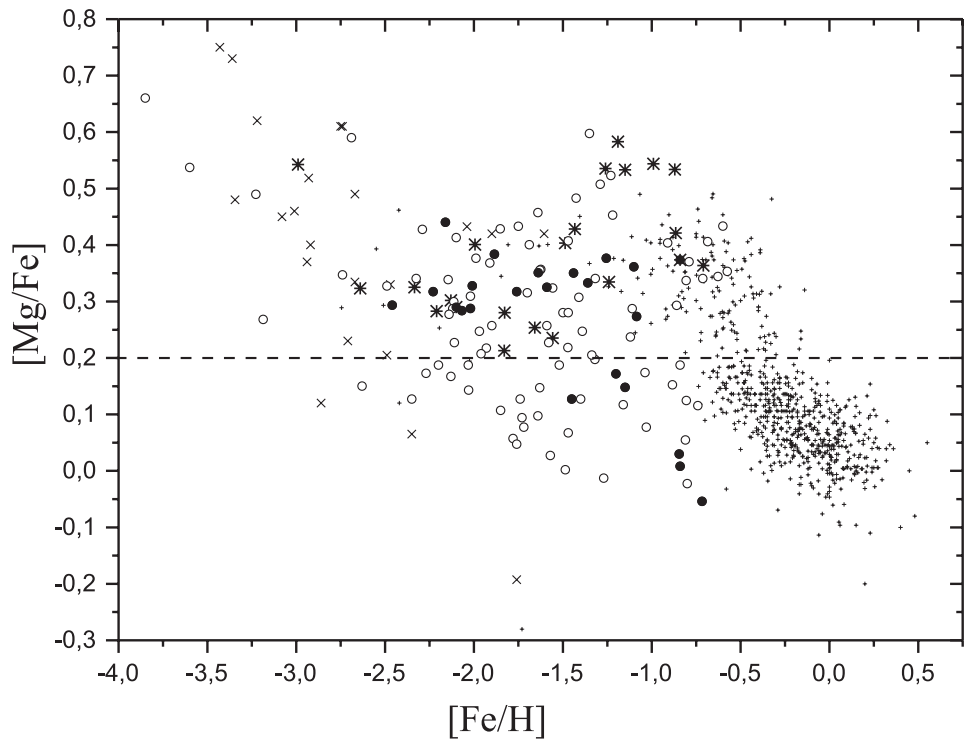}
\caption{Metallicity vs. relative magnesium abundance for all of the stars 
in the catalog. The notation is the same as that in Fig.~1; the crosses 
indicate the stars that were not identified due to the lack of radial 
velocities. The dashed line was drawn through [Mg/Fe]=0.2~dex. The filled 
circles highlight the members of the Centaurus moving group among the 
accreted stars.
}
\label{fig2}
\end{figure*}

\newpage

\begin{figure*}
\centering
\includegraphics[angle=0,width=0.80\textwidth,clip]{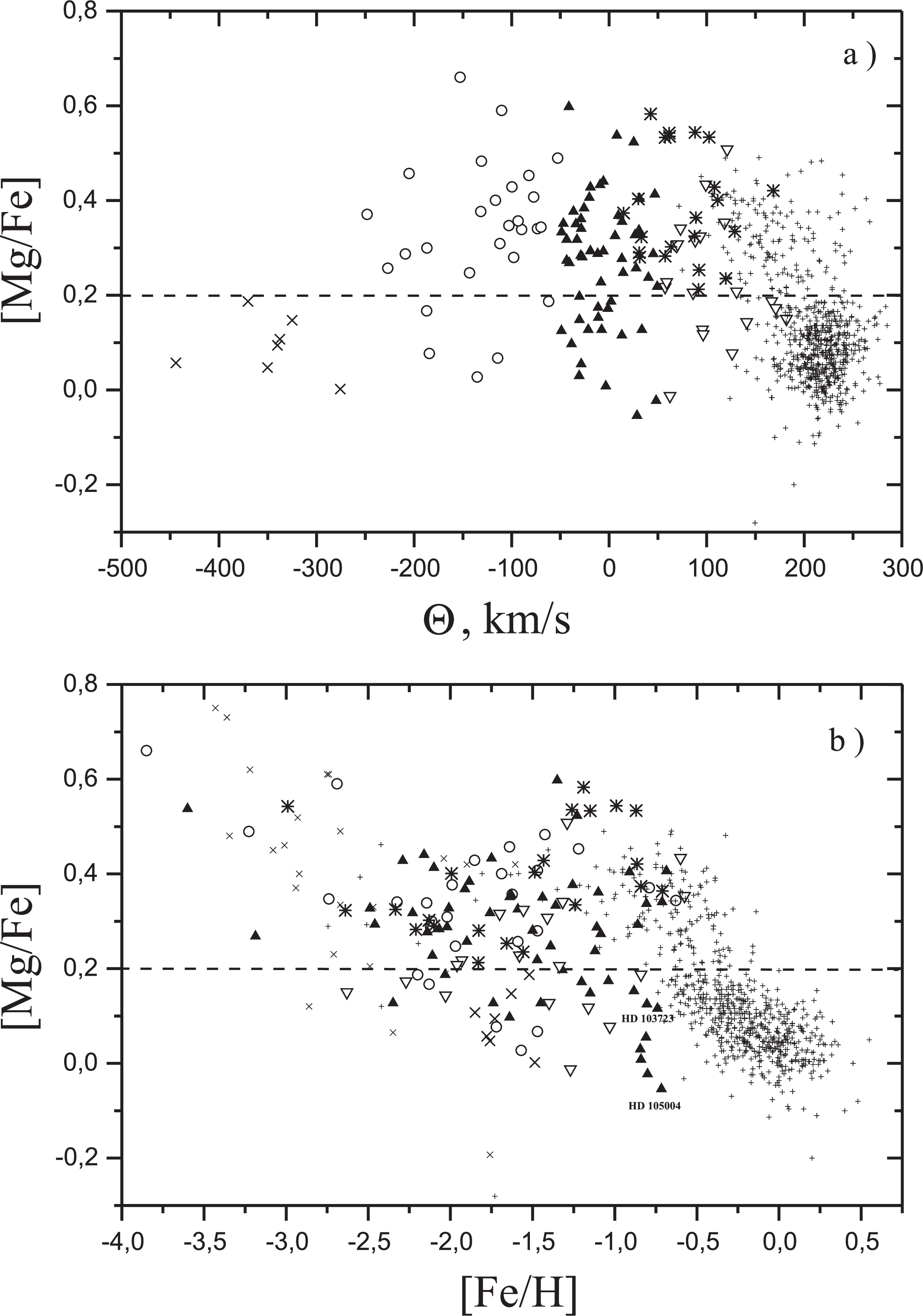}
\caption{Relative magnesium abundances vs.(a) azimuthal velocities and (b)
metallicity. The genetically related stars of the thin and thick disks 
and the protogalactic halo are denoted as in Fig.~1: the crosses, open 
circles, filled triangles with upward vortices, and open triangles with 
downward vortices represent presumably accreted stars with azimuthal 
velocities in the ranges $\Theta < -250$, $-250<\Theta <-50$, 
$-50<\Theta <+50$, and $\Theta > 50$~km\,s$^{-1}$ respectively.}
\label{fig3}
\end{figure*}

\newpage

\begin{figure*}
\centering
\includegraphics[angle=0,width=0.80\textwidth,clip]{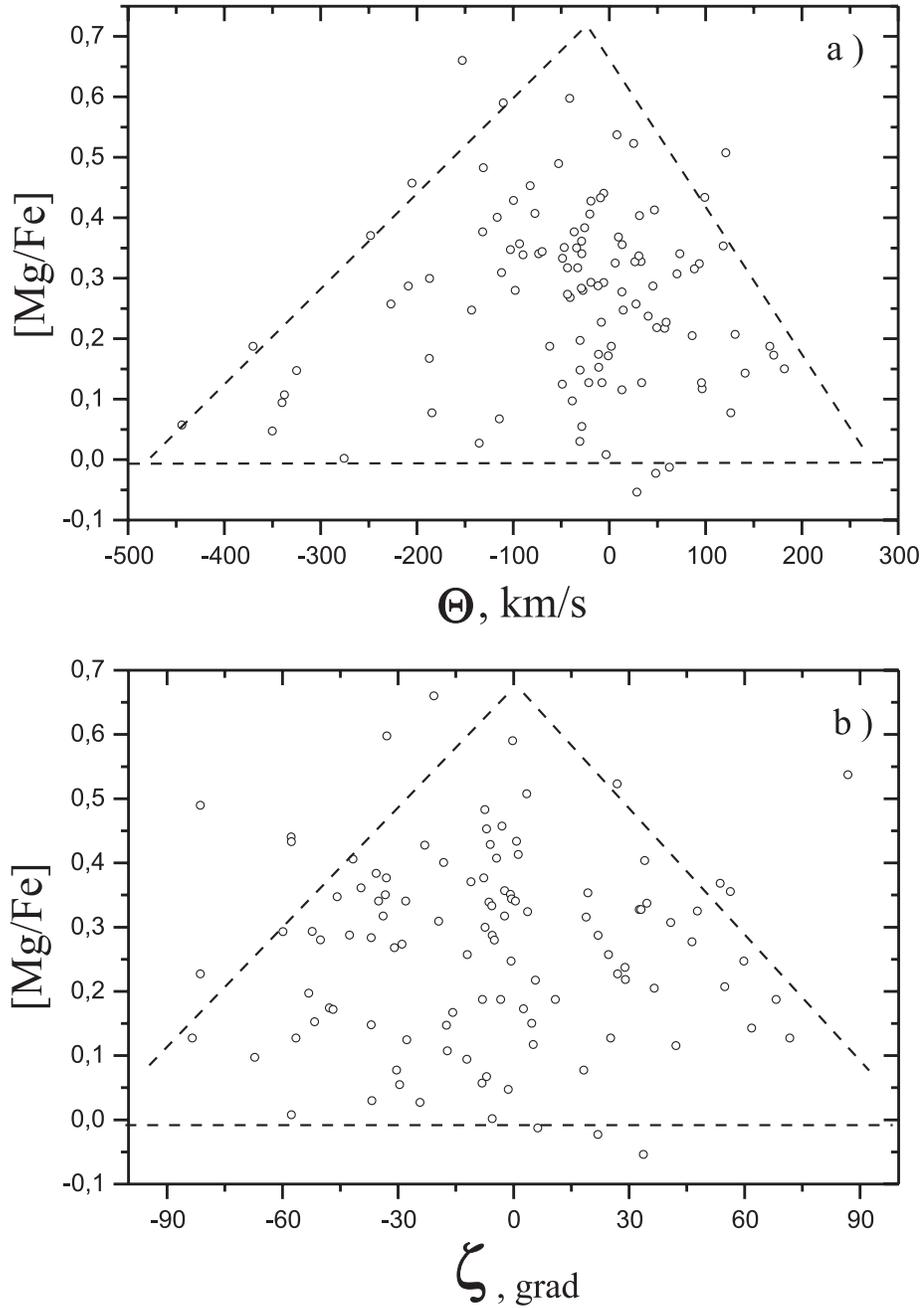}
\caption{Relative magnesium abundances in accreted stars vs. (a) 
their azimuthal velocities and (b) Galactic orbital inclinations. 
The dashed lines represent the envelopes of the points in the 
diagrams drawn by eye.}
\label{fig4}
\end{figure*}

\newpage

\begin{figure*}
\centering
\includegraphics[angle=0,width=0.99\textwidth,clip]{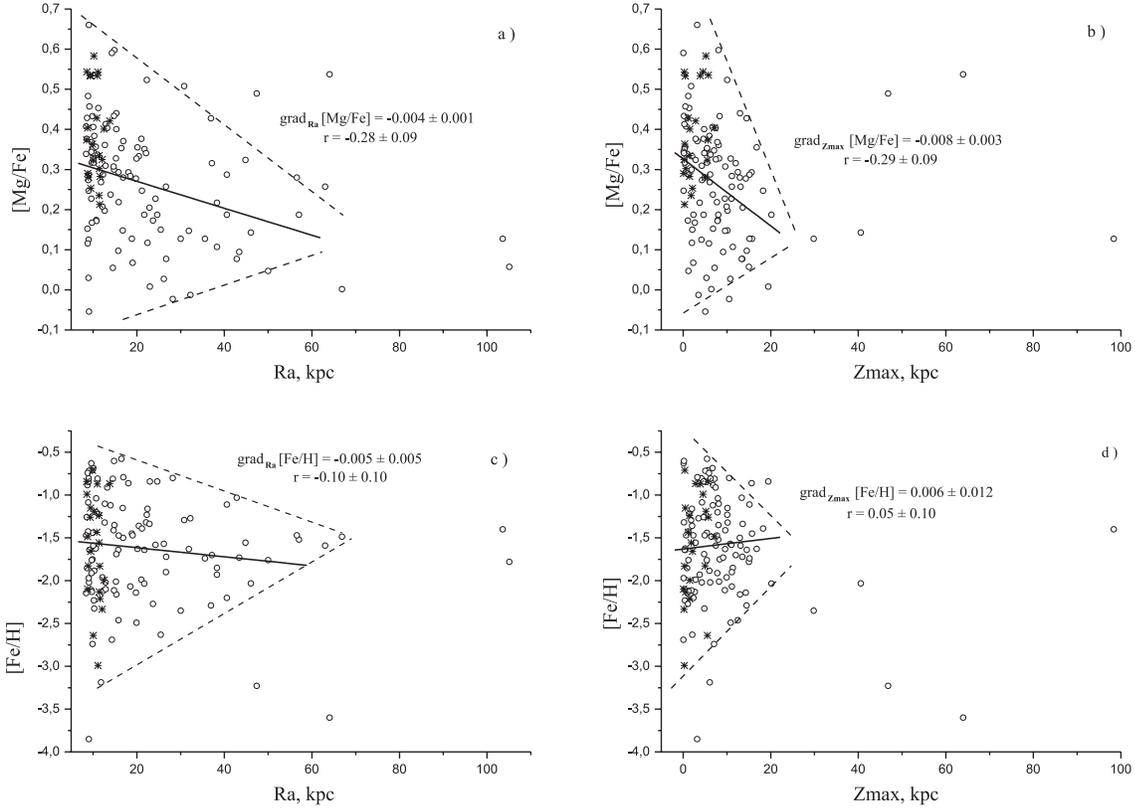}
\caption{Relative magnesium abundances (upper row) and 
metallicities (lower row) in protodisk and accreted halo 
stars vs. maximum distances of the orbital points from the 
Galactic center (a, c) and plane (b, d). The notation is the same 
as that in Fig.~1. The solid lines represent the regression lines 
for accreted halo stars. The corresponding gradients and correlation
coefficients are indicated. The dashed lines represent the lower and 
upper envelopes of the points in the diagrams drawn by eye. 
}
\label{fig5}
\end{figure*}

\newpage

\begin{figure*}
\centering
\includegraphics[angle=0,width=0.99\textwidth,clip]{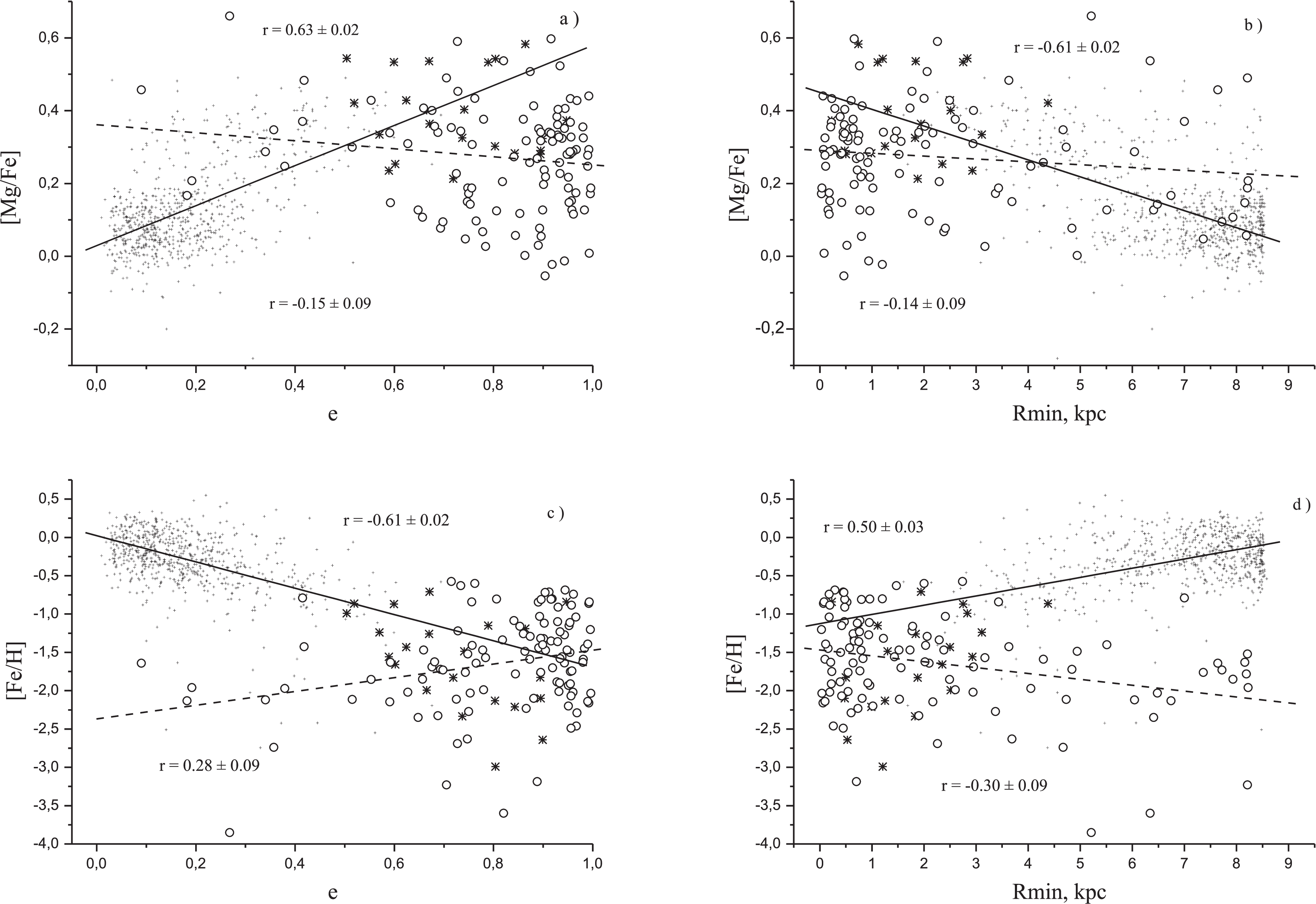}
\caption{Relative magnesium abundances (upper row) and metallicities 
(lower row) in the stars of our catalog vs. eccentricities (a, c) and 
perigalactic radii of their orbits (c,d). The notation is the same as 
that in Fig.~1. The solid and dashed lines represent the regression 
lines for genetically related stars and presumably accreted stars, 
respectively. The correlation coeffcients for genetically related 
stars and presumably accreted stars are given at the top and the bottom 
respectively.
}
\label{fig6}
\end{figure*}


\begin{thebibliography}{}

\bibitem{Abadi03}
M.G.~Abadi, M.G.~Navarro, M.~Steinmetzand, and
V.R.~Eke, Astrophys.J. {\bf 591}, 499, (2003).

\bibitem{Allen91}
C.~Allen and A.~Santillan, Rev.Mex.Astron.Astrophys.
{\bf 22}, 255 (1991).

\bibitem{Borkova00}
T.V.~Borkova and V.A.~Marsakov, Astron.Zh. {\bf 77}, 750
(2000) [Astron.Rep. {\bf 44}, 665 (2000)].

\bibitem{Borkova02}
T.V.~Borkova and V.A.~Marsakov, Bull.Spec.Astrophys.Obs.
{\bf 54}, 61 (2002).

\bibitem{Borkova04}
T.V.~Borkova and V.A.~Marsakov, Pis'ma Astron.Zh.
{\bf 30}, 173 (2004) [Astron.Lett. {\bf 30}, 148 (2004)].

\bibitem{Borkova05}
T.V.~Borkova and V.A.~Marsakov, Astron.Zh. {\bf 82}, 453
(2005) [Astron.Rep. {\bf 49}, 405 (2005)].

\bibitem{Decauwer05}
H.~Decauwer, E.~Jeh Cin, G.~Parmentier and P.~Magain,
Astron. Astrophys. {\bf 433}, 285 (2005).

\bibitem{Gilmore98}
G.~Gilmore and R.F.G.~Wise, Astrophys.J. {\bf 367}, L55 (1998).

\bibitem{Ibata98}
R.A.~Ibata and A.O.~Rasoumov, Astron. Astrophys. {\bf 336}, 
130 (1998).

\bibitem{Ibata97}
R.A.~Ibata, R.F.G.~Wyse, G.~Gilmor et al. Astron.J. 
{\bf 113}, 634 (1997).

\bibitem{Larson96}
R.B.~Larson, Astron.Soc.Pac.Conf.Ser. {\bf 92}, 241 (1996).

\bibitem{Marsakov05}
V.A.~Marsakov and T.V.~Borkova, Pis'ma Astron.Zh.
{\bf 31}, 577 (2005) [Astron.Rep. {\bf 31}, 515 (2005)].

\bibitem{Marsakov06}
V.A.~Marsakov and T.V.~Borkova, Pis'ma Astron.Zh. {\bf 32}, 419 (2006)
[Astron. Lett. {\bf 32}, 376 (2006)]
(in press).

\bibitem{Marsakov77}
V.A.~Marsakov and A.A.~Suchkov, Astron.Zh. {\bf 54}, 1232 (1977)
[Sov.Astron. {\bf 21}, 700 (1977)].

\bibitem{Mashonkina01}
L.~Mashonkina and T.~Gehren, Astron.Astrophys. {\bf 376}, 232 (2001).

\bibitem{Matteucci01}
F.~Matteucci, Nature, {\bf 414}, 253 (2001).

\bibitem{Matteucci89}
F.~Matteucci and P.~Francois, Mon.Not.R.Astron.Soc. {\bf 239}, 
885 (1989).

\bibitem{Meza05}
A.~Meza, J.F.~Navarro, M.G.~Abadi, and M.~Steinmetz, 
Mon.Not.R.Astron.Soc. {\bf 359}, 93 (2005).

\bibitem{Nomoto97}
K.~Nomoto, K.~Iwamoto, N.~Nakasato et al. Nucl.Phys.A 
{\bf 621}, 467 (1997).

\bibitem{Ratnatunga89}
K.U.~Ratnatunga, J.N.~Bahcall and S.~Casrtano, Astropys.J.
{\bf 291}, 260 (1989).

\bibitem{Thielemann96}
F.-K.~Thielemann, K.~Nomotto, and G.~Meyned, Astrophys.J.
{\bf 460}, 408 (1996).

\bibitem{Tolstoy03}
E.~Tolstoy K.A.~Venn, M.~Shetrone et al., Astron.J. {\bf 125}, 
707 (2003).

\bibitem{Tshuchiya03}
T.~Tshuchiya, D.~Dinescu, and V.I.~Korchagin, Astrophys.J.
{\bf 589}, L29 (2003).

\bibitem{Tsujimoto95}
T.~Tsujimoto, K.~Nomoto, Y.~Yoshii, et al., Mon.Not.R.Astron.Soc.
{\bf 277}, 945 (1995).

\end{thebibliography}
\end{document}